\documentclass[prd,onecolumn,nofootinbib,aps,preprintnumbers,12pt]{revtex4}

\usepackage{fouridx}
\usepackage{latexsym}
\usepackage{amsmath}
\usepackage{amssymb}
\usepackage{amsfonts}
\usepackage{graphicx}
\usepackage{hyperref}
\usepackage{dsfont}
\usepackage{verbatim}
\usepackage{enumerate}
\newcommand{\be}{\begin{equation}}
\newcommand{\ee}{\end{equation}}
\newcommand{\bea}{\begin{eqnarray}}
\newcommand{\eea}{\end{eqnarray}}

\def\tr{\mathop{\rm Tr}\nolimits}

\def\dim{\mathop{\rm dim}\nolimits}
\newcommand{\obs}{\widehat{\mathcal O}}
\newcommand{\Hilbert}{{\mathcal H}}

\newcommand{\ketbra}[2]{\left|#1\left\rangle \right\langle #2 \right|}

\newcommand{\avec}{\vec\alpha}
\newcommand{\id}{{\mathds{1}}}

\begin{document}
\begin{flushright} CALT-TH-2015-025  \end{flushright}

\title{Why Boltzmann Brains Don't Fluctuate Into Existence \\ From the De~Sitter Vacuum}

\author{Kimberly K.\ Boddy,$^a$ Sean M.\ Carroll,$^b$ and Jason Pollack$^b$\\
$^a$ Department of Physics and Astronomy, University of Hawai'i\\
$^b$ W. Burke Institute for Theoretical Physics, Calif. Institute of Technology\\
{\tt kboddy@hawaii.edu, seancarroll@gmail.com, jpollack@caltech.edu}}

\begin{abstract}
Many modern cosmological scenarios feature large volumes of spacetime in a de~Sitter vacuum phase.
Such models are said to be faced with a ``Boltzmann Brain problem'' -- the overwhelming majority of observers with fixed local conditions are random fluctuations in the de~Sitter vacuum, rather than arising via thermodynamically sensible evolution from a low-entropy past.
We argue that this worry can be straightforwardly avoided in the Many-Worlds (Everett) approach to quantum mechanics, as long as the underlying Hilbert space is infinite-dimensional.
In that case, de~Sitter settles into a truly stationary quantum vacuum state.
While there would be a nonzero probability for \emph{observing} Boltzmann-Brain-like fluctuations in such a state, ``observation'' refers to a specific kind of dynamical process that does not occur in the vacuum (which is, after all, time-independent).
Observers are necessarily out-of-equilibrium physical systems, which are absent in the vacuum.
Hence, the fact that projection operators corresponding to states with observers in them do not annihilate the vacuum does not imply that such observers actually come into existence.
The Boltzmann Brain problem is therefore much less generic than has been supposed.
(Based on a talk given by SMC at, and to appear in the proceedings of, the Philosophy of Cosmology conference in Tenerife, September 2014.)
\end{abstract}

\maketitle

\section{Introduction}\label{introduction}

The Boltzmann Brain problem \cite{Dyson:2002pf,Albrecht:2004ke,Bousso:2006xc} is a novel constraint on cosmological models.
It arises when there are thought to be very large spacetime volumes in a de~Sitter vacuum state -- empty space with a positive cosmological constant $\Lambda$.
This could apply to theories of eternal inflation and the cosmological multiverse, but it also arises in the future of our current universe, according to the popular $\Lambda$CDM cosmology.

Observers in de~Sitter are surrounded by a cosmological horizon at a distance $R= H^{-1}$, where $H=\sqrt{\Lambda/3}$ is the (fixed) Hubble parameter.
Such horizons are associated with a finite entropy $S = 3\pi/G\Lambda$ and temperature $T=H/2\pi$~\cite{Gibbons:1977mu}.
With a finite temperature and spatial volume, and an infinite amount of time, it has been suggested that we should expect quantum/thermal fluctuations into all allowed configurations.
In this context, any particular kind of anthropically interesting situation (such as an individual conscious ``brain,'' or the current macrostate of the room you are now in, or the Earth and its biosphere) will fluctuate into existence many times.
With very high probability, when we conditionalize on the appearance of some local situation, the rest of the state of the universe will be generic -- close to thermal equilibrium -- and both the past and future will be higher-entropy states.\footnote{The real problem with an eternally fluctuating universe is not that it would look very different from ours. It's that it would contain observers who see exactly what we see, but have no reason to take any of their observations as reliable indicators of external reality, since the mental impressions of those observations are likely to have randomly fluctuated into their brains.}
These features are wildly different from the universe we think we actually live in, featuring a low-entropy Big Bang state approximately 13.8 billion years ago.
Therefore, the story goes, our universe must not be one with sufficiently large de~Sitter regions to allow such fluctuations to dominate.

In this article we summarize and amplify a previous paper in which we argued that the Boltzmann Brain problem is less generic (and therefore more easily avoided) than is often supposed \cite{Boddy:2014eba}.
Our argument involves a more precise understanding of the informal notion of ``quantum fluctuations.''
This term is used in ambiguous ways when we are talking about conventional laboratory physics: it might refer to Boltzmann (thermal) fluctuations, where the microstate of the system is truly time-dependent; or measurement-induced fluctuations, where repeated observation of a quantum system returns stochastic results; or time-independent ``fluctuations'' of particles or fields in the vacuum, which are really just a poetic way of distinguishing between quantum and classical behavior.
In the de~Sitter vacuum, which is a stationary state, there are time-independent vacuum fluctuations, but there are no dynamical processes that could actually bring Boltzmann Brains (or related configurations) into existence.
Working in the Everett (Many-Worlds) formulation of quantum mechanics, we argue that the kinds of events where something may be said to ``fluctuate into existence'' are dynamical processes in which branches of the wave function decohere.
Having a nonzero \emph{amplitude} for a certain quantum event should not be directly associated with the probability that such an event will \emph{happen}; things only happen when the wave function branches into worlds in which those things occur.

Given this understanding, the Boltzmann Brain problem is avoided when horizon-sized patches of the universe evolve toward the de~Sitter vacuum state.
This is generically to be expected in the context of quantum field theory in curved spacetime, according to the cosmic no-hair theorems~\cite{Wald:1983ky,Hollands:2010pr,Marolf:2010nz}.
It would not be expected in the context of horizon complementarity in a theory with a true de~Sitter minimum; there, the whole theory is described by a finite-dimensional Hilbert space, and we should expect Poincar\'e recurrences and Boltzmann fluctuations~\cite{Stephens:1993an,Susskind:1993if,Banks:2000fe,Banks:2001yp,Parikh:2002py,Dyson:2002pf,Albrecht:2004ke,Bousso:2006xc}. 
Such theories do have a Boltzmann Brain problem.
However, if we consider a $\Lambda > 0$ false-vacuum state in a theory where there is also a $\Lambda=0$ state, the theory as a whole has an infinite-dimensional Hilbert space.
Then we would expect the false-vacuum state, considered by itself, to dissipate toward a quiescent state, free of dynamical fluctuations.
Hence, the Boltzmann Brain problem is easier to avoid than conventionally imagined.

Our argument raises an interesting issue concerning what ``really happens'' in the Everettian wave function.
We briefly discuss this issue in Section~\ref{happens}.

\section{Quantum Fluctuations in Everettian Quantum Mechanics}

The existence of Boltzmann Brain fluctuations is a rare example of a question whose answer depends sensitively on one's preferred formulation of quantum theory.
Here we investigate the issue in the context of Everettian quantum mechanics (EQM) \cite{Everett:1957hd,wallace,Sebens:2014iwa}.
The underlying ontology of EQM is extremely simple, coming down to two postulates:
\begin{enumerate}
\item The world is fully represented by quantum states $|\psi\rangle$ that are elements of a Hilbert space $\Hilbert$.
\item States evolve with time according to the Schr\"odinger equation,
\be
  \hat{H}|\psi(t)\rangle = i\partial_t|\psi(t)\rangle,
\ee
for some self-adjoint Hamiltonian operator $\hat H$.
\end{enumerate}
The challenge, of course, is matching this austere framework onto empirical reality.
In EQM, our task is to derive, rather than posit, features such as the apparent collapse of the wave function (even though the true dynamics are completely unitary) and the Born Rule for probabilities (even though the full theory is completely deterministic).
We won't delve into these issues here, but only emphasize that in EQM the quantum state and its unitary evolution are assumed to give a complete description of reality.
No other physical variables or measurement postulates are required.

Within this framework, consider a toy system such as a one-dimensional simple harmonic oscillator with potential $V(x) = \frac{1}{2}\omega^2x^2$.
Its ground state is a Gaussian wave function whose only time-dependence is an overall phase factor, 
\be
 \psi(x,t) \propto e^{-i E_0 t} e^{-E_0 x^2}.
\ee
The phase is of course physically irrelevant; one way of seeing that is to note that equivalent information is encoded in the pure-state density operator,
\be
  \rho(x,t) = |\psi(x,t)\rangle\langle\psi(x,t)|  = |\psi(x,0)\rangle\langle\psi(x,0)| ,
\ee
which is manifestly independent of time.
We will refer to such states, which of course would include any energy eigenstate of any system with a time-independent Hamiltonian, as ``stationary.''

In a stationary state, there is nothing about the isolated quantum system that is true at one time but not true at another time.
There is no sense, therefore, in which anything is dynamically fluctuating into existence.
Nevertheless, we often informally talk about ``quantum fluctuations'' in such contexts, whether we are considering a simple harmonic oscillator, an electron in its lowest-energy atomic orbital, or vacuum fluctuations in quantum field theory.
Clearly it is important to separate this casual notion of fluctuations from true time-dependent processes.

To that end, it is useful to distinguish between different concepts that are related to the informal notion of ``quantum fluctuations."
We can identify three such ideas:
\begin{itemize}
\item \textbf{Vacuum Fluctuations} are the differences in properties of a quantum and its classical analogue, and exist even in stationary states.

\thinspace

\item \textbf{Boltzmann Fluctuations} are dynamical processes that arise when the microstate of a system is time-dependent even if its coarse-grained macro\-state may not be.

\thinspace

\item \textbf{Measurement-Induced Fluctuations} are the stochastic observational outcomes resulting from the interaction of a system with a measurement device, followed by decoherence and branching.
\end{itemize}

Let us amplify these definitions a bit.
By ``vacuum fluctuations'' we mean the simple fact that quantum states, even while stationary, generally have nonzero variance for observable properties.
Given some observable $\obs$, we expect expectation values in a state $|\psi\rangle$ to satisfy $\langle\obs^2\rangle_\psi > \langle\obs\rangle_\psi^2$.
The fact that the position of the harmonic oscillator is not localized to the origin in its ground state is a consequence of this kind of fluctuation.
Other manifestations include the Casimir effect, the Lamb shift, and radiative corrections due to virtual particles in quantum field theory.
Nothing in our analysis denies the existence of these kinds of fluctuation; we are merely pointing out that they are non-dynamical, and therefore not associated with anything literally fluctuating into existence.

This is in contrast with ``Boltzmann fluctuations,'' which are true dynamical processes.
In classical statistical mechanics, we might have a system in equilibrium described by a canonical ensemble, where macroscopic quantities such as temperature and density are time-independent.
Nevertheless, any particular realization of such a system is represented by a microstate with true time-dependence; the molecules in a box of gas are moving around, even in equilibrium.
From a Boltzmannian perspective, we coarse-grain phase space into macrostates, and associate to each microstate and entropy $S= k_B \log \Omega$, where $\Omega$ is the volume of the macrostate in which the microstate lives.
We then expect rare fluctuations into lower-entropy states, with a probability that scales as $P(\Delta S) \sim e^{-\Delta S}$, where $\Delta S$ is the decrease in entropy.
Such Boltzmann fluctuations are \emph{not} expected to occur in stationary quantum states, where there is no microscopic property that is actually fluctuating beneath the surface (at least in EQM).

This can even be true in ``thermal'' states in quantum mechanics.
Consider a composite system $AB$, with weak coupling between the two factors, and $A$ much smaller than $B$.
When the composite system is in a stationary pure state $|\psi\rangle$, we expect the reduced density matrix of the subsystem to look thermal,
\begin{equation}
  \rho_A = \tr_B |\psi\rangle\langle\psi| \sim \exp(-\beta \hat{H}_A) = \sum_n e^{-\beta E_n} \ketbra{E_n}{E_n} ,
\end{equation}
where $\hat{H}_A$ is the Hamiltonian for $A$, $\beta$ is the inverse temperature, and the states $|E_n\rangle$ are energy eigenstates of $\hat{H}_A$.
Despite the thermal nature of this density operator, it is strictly time-independent, and there are no dynamical fluctuations.

Finally, we have measurement-induced fluctuations: processes in which we repeatedly measure a quantum system and obtain ``fluctuating'' results.
In EQM, the measurement process consists of unitary dynamics creating entanglement between the observed system and a macroscopic apparatus, followed by decoherence and branching.
We can decompose Hilbert space into factors representing the system, the apparatus (a macroscopic configuration that may or may not include observing agents), and the environment (a large set of degrees of freedom that we don't keep track of):
\begin{equation}
  \Hilbert = \Hilbert_S \otimes \Hilbert_A \otimes \Hilbert_E \ .
  \label{hilbertsae}
\end{equation}
We assume that the apparatus begins in a specific ``ready'' state, and both the apparatus and environment are initially unentangled with the system to be observed.\footnote{The justification for these assumptions can ultimately be traced to the low-entropy state of the early universe.}
For simplicity, imagine that the system is a single qubit in a superposition of up and down.
Unitary evolution then proceeds as follows:
\begin{align}
|\Psi\rangle & = (|+\rangle_S + |-\rangle_S)|a_0\rangle_A |e_0\rangle_E \\
&\rightarrow (|+\rangle_S|a_+\rangle_A + |-\rangle_S|a_-\rangle_A) |e_0\rangle_E \\
&\rightarrow |+\rangle_S|a_+\rangle_A|e_+\rangle_E + |-\rangle_S|a_-\rangle_A |e_-\rangle_E .
\end{align}
The first line represents the system in some superposition of $|+\rangle$ and $|-\rangle$, while the apparatus and environment are unentangled with it.
In the second line (pre-measurement), the apparatus has interacted with the system; its readout value ``+'' is entangled with the $+$ state of the qubit, and likewise for ``$-$.''
In the final line, the apparatus has become entangled with the environment.
This is the decoherence step; generically, the environment states will quickly become very close to orthogonal, $\langle e_+|e_-\rangle \approx 0$, after which the two branches of the wave function will evolve essentially independently.
If we imagine setting up a system in some stationary state, performing a measurement, re-setting it, and repeating the process, the resulting record of measurement readouts will form a stochastic series of quantities obeying the statistics of the Born Rule.
This is the kind of ``fluctuation'' that would arise from the measurement process.

\begin{figure}[t]
  \begin{center}
    \includegraphics[width=0.92\textwidth]{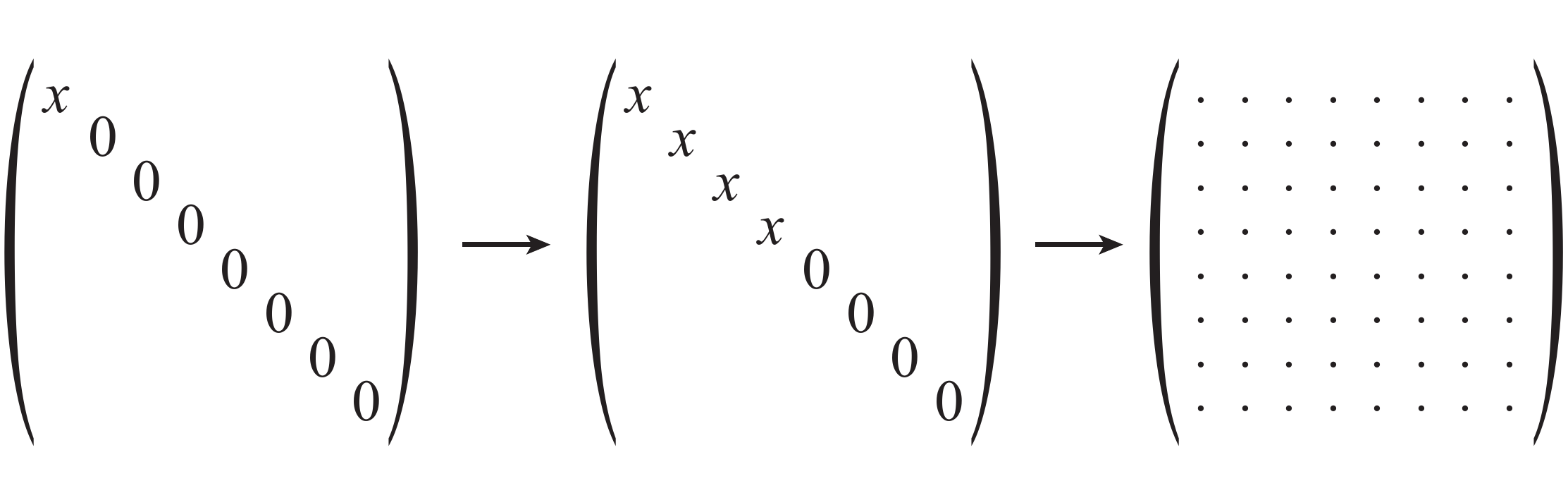}
  \end{center}
  \caption[Schematic evolution of a reduced density matrix in the pointer basis]{
    Schematic evolution of a reduced density matrix in the pointer basis.
    The density matrix on the left represents a low-entropy situation, where only a few states are represented in the wave function.
    There are no off-diagonal terms, since the pointer states rapidly decohere.
    The second matrix represents the situation after the wave function has branched a few times.
    In the third matrix, the system has reached equilibrium; the density matrix would be diagonal in an energy eigenbasis, but in the pointer basis, decoherence has disappeared and the off-diagonal terms are nonzero.}
\label{equilibration-matrices}
\end{figure}

There are several things to note about this description of the measurement process in EQM.
First, the reduced density matrix $\rho_{SA} = \tr_E |\Psi\rangle\langle \Psi|$ obtained by tracing over the environment is diagonal in a very specific basis, the ``pointer basis'' for the apparatus~\cite{Zurek:1981xq,Zurek:1993ptp,Zurek:1998ji,Zurek:2003rmp,Schlosshauer:2003zy}. 
The pointer states making up this basis are those that are robust with respect to continual monitoring by the environment; in realistic situations, this amounts to states that have a definite spatial profile (such as the pointer on a measuring device indicating some specific result).
Second, branching is necessarily an out-of-equilibrium process.
The initial state is highly non-generic; one way of seeing this is that the reduced density matrix has an initially low entropy $S_{SA} = \tr \rho_{SA} \log \rho_{SA}$.
Third, this entropy increases during the measurement process, in accordance with the thermodynamic arrow of time.
Given sufficient time to evolve, the system will approach equilibrium and the entropy will be maximal.
At this point the density matrix will no longer be diagonal in the pointer basis (it will be thermal, and hence diagonal in the energy eigenbasis).
This process is portrayed in Figure~\ref{equilibration-matrices}.
Needless to say, none of these features -- a special, out-of-equilibrium initial state, in which entropy increases as the system becomes increasingly entangled with the environment over time -- apply to isolated stationary systems.

The relationship of fluctuations and observations is worth emphasizing. 
Consider again the one-dimensional harmonic oscillator.
We can imagine constructing a projection operator onto the positive values of the coordinate,
\be
  \hat{P}_+ = \int_{x>0} dx\, |x\rangle\langle x|.
\ee
Now in some state $|\psi\rangle$, we can consider the quantity
\be
  p_+=\langle \psi|\hat{P}_+|\psi\rangle .
  \label{posproj}
\ee
In conventional laboratory settings, it makes sense to think of this as ``the probability that the particle is at $x>0$.'' 
But that's not strictly correct in EQM.
There is no such thing as ``where the particle is''; rather, the state of the particle is described by its entire wave function.
The quantity $p_+$ is the probability that we would \emph{observe} the particle at $x>0$ were we to measure its position.
Quantum variables are not equivalent to classical stochastic variables. 
They may behave similarly when measured repeatedly over time, in which case it is sensible to identify the nonzero variance of a quantum-mechanical observable with the physical fluctuations of a classical variable, but the state in EQM is simply the wave function, not the collection of possible measurement outcomes.

\section{Boltzmann Brains and De Sitter Space}

With this setup in mind, the application to de~Sitter space is straightforward.
As mentioned in the Introduction, observers in de~Sitter are surrounded by a horizon with a finite entropy.
In the vacuum, the quantum state in any horizon patch is given by a time-independent thermal density matrix, 
\be
  \rho_{\textrm{patch}} \propto e^{-\beta \hat{H}},
\ee
where the Hamiltonian is the static Hamiltonian for the fields in that patch and $\beta \propto 1/\sqrt{\Lambda}$.

According to the analysis in the previous section, this kind of thermal state does \emph{not} exhibit dynamical fluctuations of any sort, including into Boltzmann Brains.
It is a stationary state, so there is no time-dependence in any process.
In particular, the entropy is maximal for the thermal density matrix, so there are no processes corresponding to decoherence and branching.\footnote{The idea that quantum fluctuations during inflation are responsible for the density perturbations in our current universe is unaffected by this reasoning. During inflation the state is nearly stationary, with non-dynamical vacuum fluctuations as defined above; branching and decoherence occur when the entropy ultimately increases, for example during reheating.}
There may be nonzero overlap between some state $|\textrm{brain}\rangle$ and the de~Sitter vacuum, but there is no dynamics that brings that state into existence on a decoherent branch of the wave function.
Indeed, one way of establishing the thermal nature of the state is to notice that a particle detector placed in de~Sitter space will come to equilibrium and then stop evolving~\cite{Spradlin:2001pw}.
Therefore, Boltzmann Brains do not fluctuate into existence in such a state, and should not be counted among observers in the cosmological multiverse.

It is useful to contrast this situation with that of a black hole in a Minkowski background.
There, as in de~Sitter space, we have a horizon with a nonzero temperature and finite entropy.
However, real-world black holes are not stationary states.
They evaporate by emitting radiation, and the entropy increases along the way. 
A particle detector placed in orbit around a black hole will not simply come to equilibrium and stop evolving; it will detect particles being emitted from the direction of the hole, with a gradually increasing temperature as the hole shrinks.
This is a very different situation than the equilibrium de~Sitter vacuum.

\begin{figure}[t]
  \begin{center}
    \includegraphics[width=0.32\textwidth]{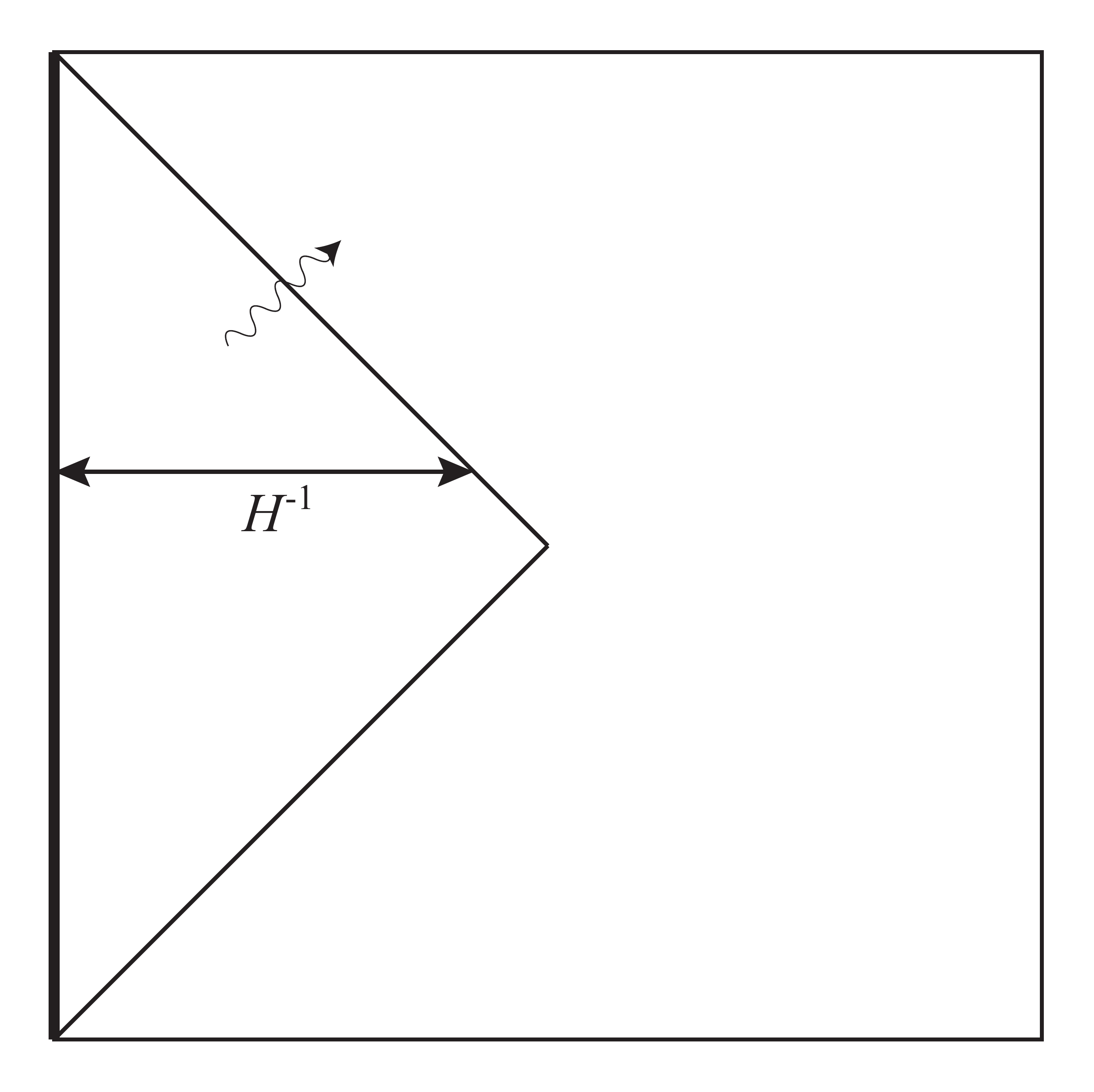}
    \includegraphics[width=0.32\textwidth]{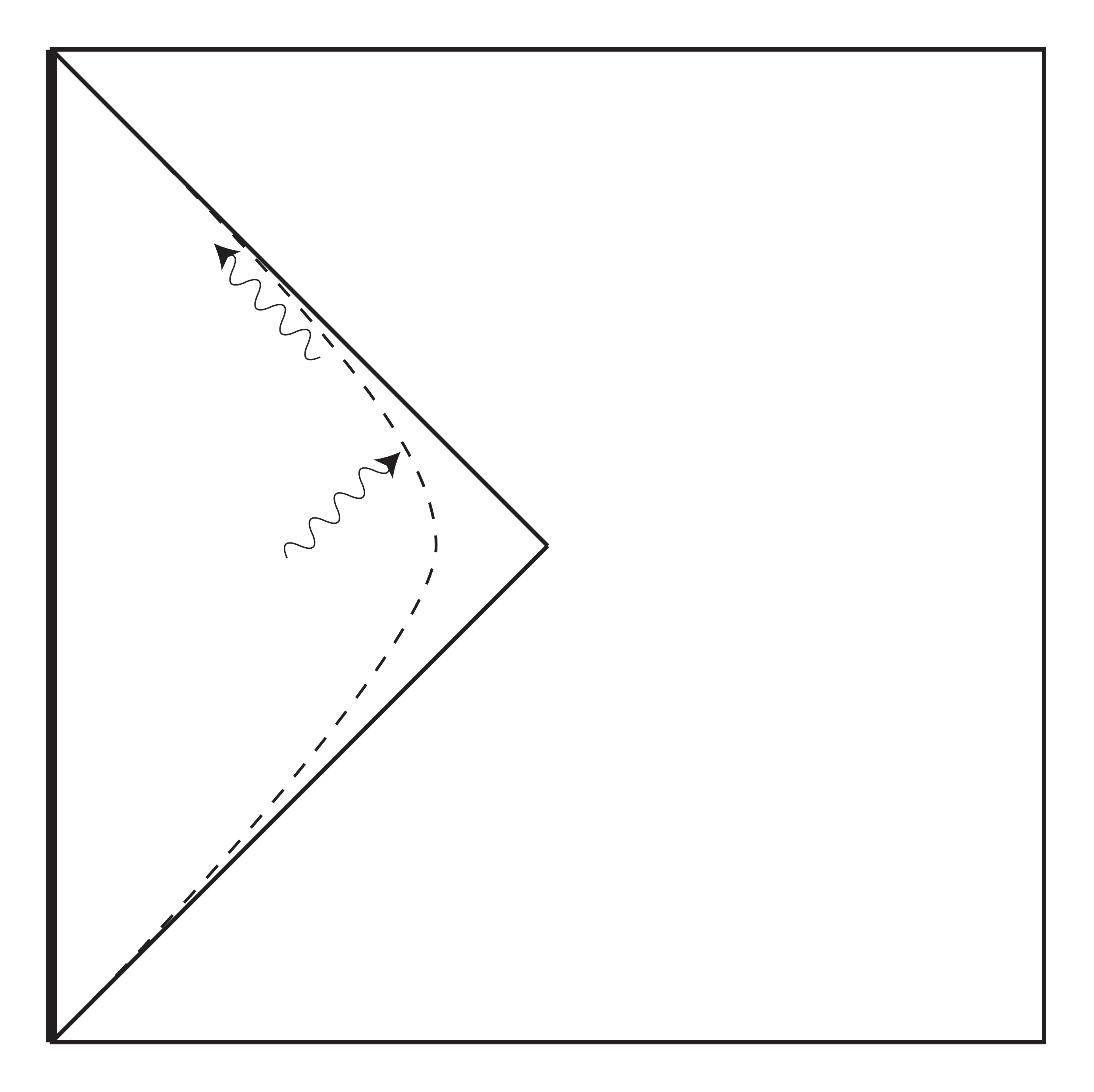}
    \includegraphics[width=0.32\textwidth]{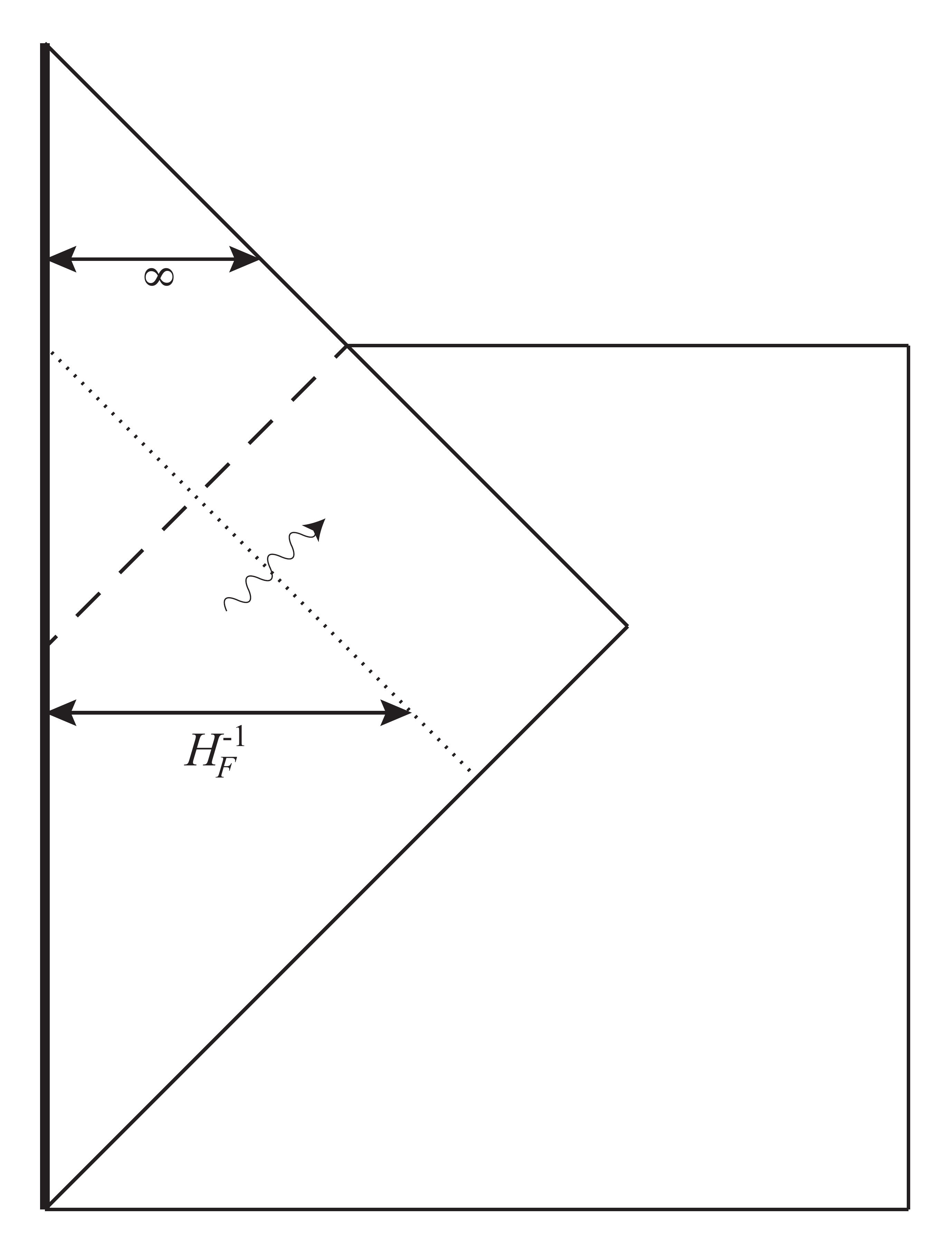}
  \end{center}
  \caption[Conformal diagrams for de~Sitter space]{
    Conformal diagrams for de~Sitter space in different scenarios.
    We consider an observer at the north pole, represented by the line on the left boundary and their causal diamond (solid triangle).
    The wavy line represents excitations of the vacuum approaching the horizon.
    In QFT in curved spacetime, portrayed on the left, the excitation exits and the state inside the diamond approaches the vacuum, in accordance with the cosmic no-hair theorems.
    In contrast, horizon complementarity implies that excitations are effectively absorbed at the stretched horizon (dashed curve just inside the true horizon) and eventually return to the bulk, as shown in the middle diagram.
    The third diagram portrays the situation when the de~Sitter minimum is a false vacuum, and the full theory contains a state with $\Lambda=0$; the upper triangle represents nucleation of a bubble of this Minkowski vacuum.
    In that case, excitations can leave the apparent horizon of the false vacuum while remaining inside the true horizon; we then expect there to be no dynamical Boltzmann fluctuations.}
\label{conformal-fig}
\end{figure}

It remains to determine whether the stationary vacuum state is actually attained in the course of cosmological evolution.
There are classical and quantum versions of the cosmic no-hair theorem \cite{Wald:1983ky,Hollands:2010pr,Marolf:2010nz}.
Classically, the spacetime geometry of each horizon-sized patch of a universe with $\Lambda > 0$ asymptotically approaches that of de~Sitter space, as long as it does not contain a perturbation so large that it collapses to a singularity.
In the context of quantum field theory in curved spacetime, analogous results show that each patch approaches the de~Sitter vacuum state.
Intuitively, this behavior can be thought of as excitations leaving the horizon and not coming back, as portrayed in the first part of Figure~\ref{conformal-fig}.
The timescale for this process is parametrically set by the Hubble time, and will generally be enormously faster than the rate of Boltzmann fluctuations in states that have not quite reached the vacuum.
Hence, if we think of conventional $\Lambda$CDM cosmology in terms of semiclassical quantum gravity, it seems reasonable to suppose that the model does not suffer from a Boltzmann Brain problem.

The situation is somewhat different if we take quantum gravity into account.
In this case we are lacking a fully well-defined theory, and any statements we make must have a conjectural aspect.
A clue, however, is provided the idea of horizon complementarity~\cite{Stephens:1993an,Susskind:1993if,Banks:2000fe,Banks:2001yp,Parikh:2002py}.
According to this idea, we should only attribute a local spacetime description to regions on one side of any horizon at a time, rather than globally.
For example, we could describe the spacetime outside of a black hole, or as seen by an observer who has fallen into the black hole, but shouldn't use both descriptions simultaneously; the rest of the quantum state can be thought of as living on a timelike ``stretched horizon'' just outside the null horizon itself.
Applying this philosophy to de~Sitter space leads to the idea that the whole theory should be thought of as that of a single horizon patch, with everything normally associated with the rest of the universe actually encoded on the stretched cosmological horizon.
Since the patch has a finite entropy (approximately $10^{123}$ for the measured value of $\Lambda$), the corresponding quantum theory is (plausibly) finite-dimensional, with $\dim\Hilbert = e^S$.

Hence, applying horizon complementarity to a universe with a single true de~Sitter vacuum, the intuitive picture behind the cosmic no-hair theorem no longer applies.
There is not outside world for perturbations to escape to; rather, they are absorbed by the stretched horizon, and will eventually be emitted back into the bulk spacetime, as shown in the middle part of Figure~\ref{conformal-fig}.
This is consistent with our expectations for a quantum theory on a finite-dimensional Hilbert space, which should exhibit fluctuations and Poincar\'e recurrences.
This was the case originally considered by Dyson, Kleban, and Susskind~\cite{Dyson:2002pf,Albrecht:2004ke} in their examination of what is now known as the Boltzmann Brain problem.
Nothing in our analysis changes this expectation; if Hilbert space is finite-dimensional and time evolves eternally, it is natural to expect that the Boltzmann Brain problem is real.
(Though see \cite{Albrecht:2014eaa} for one attempt at escaping this conclusion.)

The situation changes if the de~Sitter vacuum state is only metastable, and is embedded in a larger theory with a $\Lambda=0$ minimum.
In that case the underlying quantum theory will be infinite-dimensional, since the entropy of Minkowski space is infinite.
In a semiclassical solution based on the de~Sitter vacuum, the dynamics will not be completely unitary, since there will be interactions (such as bubble nucleations) connecting different sectors of the theory.
Although a full understanding is lacking, intuitively we expect the dynamics in such states to be dissipative, much as higher-energy excitations of metastable minima decay away faster in ordinary quantum mechanics.
The Poincar\'e recurrence time is infinite, so there is no necessity for Boltzmann fluctuations or recurrences.
The spacetime viewpoint relevant to this case is portrayed in the third panel of Figure~\ref{conformal-fig}.
The existence of Coleman-De~Luccia transitions to the $\Lambda = 0$ vacua permits the true horizon size to be larger than the de~Sitter radius, so perturbations can appear to leave the horizon and never return, even under complementarity.

The complete picture we suggest is therefore straightforward.
If we are dealing with de~Sitter vacua in a theory with an infinite-dimensional Hilbert space, we expect horizon patches to evolve to a stationary quantum vacuum state, and there to be no dynamical fluctuations, and the Boltzmann Brain problem is avoided.
This applies to QFT in curved spacetime, or to complementarity in the presence of $\Lambda = 0$ vacua.
If, on the other hand, the Hilbert space is finite, fluctuations are very natural, and the Boltzmann Brain problem is potentially very real.

\section{What Happens In the Wave Function?}\label{happens}

We have been careful to distinguish between vacuum fluctuations in a quantum state, which can be present even if the state is stationary, and true dynamical processes, such as Boltzmann and measurement-induced (branching) fluctuations.
One may ask, however, whether our interpretation of stationary states in EQM is the right one.
More specifically: is it potentially legitimate to think of a stationary quantum state as a superposition of many time-dependent states?
This is a particular aspect of a seemingly broader question: what ``happens'' inside the quantum wave function?

One way to address this question is by using the decoherent (or consistent) histories formalism~\cite{Griffiths:1984rx,Omnes:1992ag,GellMann:1992kh,Hartle:1992bf,Halliwell:1994uz,Hartle:1994na}.
This approach allows us to ask when two possible histories of a quantum system decohere from each other and can be assigned probabilities.
We might want to say that an event (such as a fluctuation into a Boltzmann Brain) ``happens" in the wave function if that event occurs as part of a history that decoheres from other histories in some consistent set.
(Though we will argue that, in fact, this criterion is too forgiving.)

Consider a closed system described by a density operator $\rho(t_0)$ at an initial time $t_0$.
We want to consider possible coarse-grained histories of the system, described by sequences of projection operators $\{\hat{P}_\alpha\}$.
These operators partition the state of the system at some time into mutually exclusive alternatives and obey
\begin{equation}
  \sum_\alpha \hat{P}_\alpha = \id \ , \qquad
  \hat{P}_\alpha \hat{P}_\beta = \delta_{\alpha\beta}\hat{P}_\alpha \ .
  \label{projectors}
\end{equation}
A history is described by a sequence of such alternatives, given by a sequence of projectors at specified times, $\{\hat{P}^{(1)}_{\avec_1}(t_1), \ldots \hat{P}^{(n)}_{\avec_n}(t_n)\}$.
At each time $t_i$, we have a distinct set of projectors $\hat{P}^{(i)}_\alpha$, and the particular history is described by a vector of specific projectors labeled by $\avec$.
The decoherence functional of two histories $\avec$ and $\avec'$ is
\begin{equation}
  D(\avec, \avec') = \tr[\hat{P}^{(n)}_{\avec_n}(t_n) \cdots
    \hat{P}^{(1)}_{\avec_1}(t_1) \rho(t_0) \hat{P}^{(1)}_{\avec_1'}(t_1) \cdots
    \hat{P}^{(n)}_{\avec_n'}(t_n)] \ ,
  \label{decoherencefunctional0}
\end{equation}
where the trace is taken over the complete Hilbert space.
If the decoherence functional vanishes for two histories, we say that those histories are consistent or decoherent, and they can be treated according to the rules of classical probability theory.

Following a suggestion by Lloyd~\cite{lloyd}, we can apply the decoherent histories formalism to a simple harmonic oscillator in its ground state.
One choice of projectors are those given by the energy eigenstates $|E_n\rangle$ themselves,
\be
  \hat{P}_n = |E_n\rangle\langle E_n|.
\ee
It is easy to check that the corresponding histories trivially decohere.
This simply reflects the fact that the system begins in an energy eigenstate and stays there.

But we are free to consider other sets of projectors as well.
Let us restrict our attention to an $N$-dimensional subspace of the infinite-dimensional oscillator Hilbert space, consisting of the span of the energy eigenstates $|E_n\rangle$ with $n$ between $0$ and $N-1$.
Then Lloyd~\cite{lloyd} points out that we can define phase states
\be
  |\phi_m\rangle = \frac{1}{\sqrt{N}}\sum_{n=0}^{N-1} e^{2\pi i mn/N} |E_n\rangle.
  \label{phasestates}
\ee
These have the property that they evolve into each other after timesteps $\Delta t = 2\pi/N\omega$,
\be
e^{-iH \Delta t}|\phi_m\rangle = |\phi_{m+1}\rangle.
\ee
Now we can consider histories defined by the phase projectors
\be
  \tilde{P}_m = |\phi_m\rangle\langle \phi_m|,
\ee
evaluated at times separated by $\Delta t$.
These histories, it is again simple to check, \emph{also} mutually decohere with each other (though, of course, not with the original energy-eigenstate histories).
Each such history describes a time-dependent system, whose phase rotates around, analogous to a classical oscillator rocking back and forth in its potential.\footnote{It was not necessary to carefully choose the phase states. In the decoherent histories formalism, we have the freedom to choose projection operators separately at each time. Given some initial projectors, we can always define projectors at later times by simply evolving them forward by an appropriate amount; the resulting histories will decohere. We thank Mark Srednicki for pointing this out.}

We therefore have two (and actually, many more) sets of histories, which decohere within the sets, but are mutually inconsistent with each other.
In some sets there is no time-dependence, while in others there is.
In the stationary thermal state of a de~Sitter horizon patch, there is no obstacle in principle to defining a set of decoherent histories with the properties that some of them describe Boltzmann Brains fluctuating into existence.
On the other hand, we are not \emph{forced} to consider such histories; there are also consistent sets in which the states remain perfectly stationary.

This situation raises a fundamental puzzle.
When we are doing multiverse cosmology, we often want to ask what is seen by observers satisfying certain criteria (which may be as general as ``all intelligent observers'' or as specific as ``observers in precisely defined local conditions").
To answer that question, we want to know whether an amplitude corresponding to such an observer is actually physically realized in the quantum state of the universe.
The decoherent histories formalism seems to give an ambiguous answer to the question: the number of observers who physically appear in the universe depends on the projection operators we choose to define our coarse-grained histories.
This seems to introduce an unacceptably subjective element into a purportedly objective calculation.
(A closely related problem has been emphasized by Kent~\cite{Kent:1996mr}.)

Our own conclusion from this analysis is simple: the existence of decoherent histories describing certain dynamical processes is \emph{not sufficient} to conclude that those processes ``really happen.''
Note that something somewhat stronger is going on in the standard description of branching and decoherence in EQM.
There, the explicit factorization of Hilbert space into system+apparatus and environment directly implies a certain appropriate coarse-graining for the macroscopic variables (by tracing over the environment).
Of course, there is arguably a subjective element in how we define the environment in the first place.
That choice, however, relies on physical properties of the theory, in particular the specific Hamiltonian and its low-energy excitations around some particular background state.
There have been suggestions that the decoherence properties of realistic systems can be defined objectively, by demanding that records of the macroscopic configuration be stored redundantly in the environment~\cite{Riedel:2013uoa}.

We conjecture, at least tentatively, that the right way to think about observers fluctuating into existence in quantum cosmology is to define an objective division of the variables into ``macroscopic system'' and ``environment,'' based on the physical properties of the system under consideration, and to look for true branching events where the reduced density matrix of the system decoheres in the pointer basis.\footnote{A possible alternative strategy is to look for histories that obey classical equations of motion, as in \cite{GellMann:1992kh,GellMann:2006uj}. Such an approach seems unable to resolve the ambiguity presented by the simple harmonic oscillator.}
Work clearly remains to be done in order to turn this idea into a well-defined program, as well as to justify why such a procedure is the appropriate one.
In this context, it is useful to keep in mind that Boltzmann Brains are a difficulty, not a desirable feature, of a given cosmological model. 
We suggest that the analysis presented here should at the very least shift the burden of proof onto those who believe that Boltzmann Brains are a generic problem.

\section*{Acknowledgements}

SMC would like to thank the organizers of the Philosophy of Cosmology workshop in Tenerife -- Joe Silk, Simon Saunders, Khalil Chamcham, John Barrow, Barry Loewer, and David Albert -- for putting together such a unique and stimulating meeting.
We are grateful to Jim Hartle, Seth Lloyd, and Mark Srednicki for conversations on the issues discussed in Section~\ref{happens}.
This research is funded in part by the Walter Burke Institute for Theoretical Physics at Caltech, by DOE grant DE-SC0011632, and by the Gordon and Betty Moore Foundation through Grant 776 to the Caltech Moore Center for Theoretical Cosmology and Physics.

\bibliography{tenerife-bib}
\bibliographystyle{utphys}

\end{document}